\documentstyle[preprint,aps,psfig,epsfig]{revtex}

\tighten
\begin{document}
\def\eqn#1{Eq.$\,$#1}
\draft
\preprint{}
\title{Analytical results for random walks in the presence of disorder and
traps}

\author{Cl\'ement Sire}
\address{Laboratoire de Physique Quantique (UMR C5626 du CNRS),
Universit\'e Paul Sabatier
31062 Toulouse Cedex, France (clement@irsamc2.ups-tlse.fr)}

\maketitle  \begin{abstract} In this paper, we study the dynamics of a random
walker diffusing on a disordered one-dimensional lattice with random trappings.
The distribution of escape probabilities is computed exactly for any strength
of the disorder. These probabilities do not display any multifractal
properties contrary to previous numerical claims. The explanation for this
apparent multifractal behavior is given, and our conclusion are supported by
numerical calculations. These exact results are exploited
to compute the large time asymptotics of the survival probability (or the
density) which is
found to decay as $\exp [-Ct^{1/3}\log^{2/3}(t)]$. An exact lower bound for
the density is found to decay in  a similar way.
\end{abstract}
\vspace{1cm}

\pacs{PACS numbers: 05.40.+j, 05.60.+w, 02.50.Ey, 61.43.Hv}

\narrowtext

\newpage

\section{Introduction}

The dynamics of the survival probability of particles diffusing in the
presence of traps is a rich problem which has been widely discussed in the
physical and mathematical literature within the last two decades
\cite{donsker,revpstat,pstat,havlin,pdiff}.

The simplest system is that of diffusing particles in the presence
of perfect static traps \cite{donsker,revpstat,pstat,havlin}. This problem
(that we will call the Donsker-Varadhan problem) has been solved using very
different technics. The main results is that the density does not decay
exponentially (as a simple mean-field argument would predict) but as,
\begin{equation}
n(t)\sim\exp \left[-C_d c^{\frac{2}{d+2}} t^{\frac{d}{d+2}}\right],
\label{don}
\end{equation}
where $c$ is the trapping site density. The physical interpretation in $d$
dimension is that the process is dominated by particles standing in very large
trap-free regions of linear size $L$ (these regions have a probability of
order $\exp(-cL^d)$). In such a region, the density decays as $\exp(-t/L^2)$. A
saddle-point argument then leads to the result of  \eqn{(\ref{don})}, 
with the relevant regions being of typical size $L\sim (t/c)^{\frac{1}{d+2}}$,
at time $t$.

In another class of models \cite{pdiff}, the traps are allowed to move. When
these traps undergo free diffusion the density of particles decays as,
\begin{equation}
n(t)\sim\exp \left[-C_d c t^{\frac{d}{2}}\right],
\end{equation}
for $d<2$, and decays 
exponentially for $d>2$.
This result holds in the case of static or diffusing particles \cite{pdiff}.

It would be interesting to introduce the effects of hopping disorder on the
trapping process. Even without trapping, quenched disorder in the particle
hopping probabilities is known to have very important effects on the diffusion
and first return properties \cite{alex,bouchaud,sinai}. In the case of
symmetric hopping probabilities ($w_{i,i+1}=w_{i+1,i})$ \cite{alex,bouchaud},
anomalous diffusion is observed, with an exponent depending  on the
properties of the disorder. In the generic non symmetric case (see
\cite{bouchaud} for a more precise criterion), as  in the Sinai model
\cite{sinai} where a particle diffuses in a random (Brownian) potential, the
diffusion is dramatically suppressed, the particle being effectively trapped
in deeper and deeper valleys of the potential as time goes on.

In the present article, we study the dynamics of particles diffusing in a
symmetric or non symmetric disorder, in the presence of a random finite
trapping probability at each site.

\section{Model and known results}

Consider a particle moving on a one-dimensional lattice with random barriers
(or hopping probabilities) and random trapping probabilities. More precisely,
a particle at site $i$ has a probability $w_{i,i+1}<1/2$ (respectively 
$w_{i,i-1}<1/2$) to hop on site $i+1$ (respectively $i-1$), and a probability
$(1-\gamma)(1-w_{i,i-1}-w_{i,i+1})$ to disappear ($\gamma<1$). With residual
probability $\gamma(1-w_{i,i-1}-w_{i,i+1})$, it just stays on site $i$.  The
hopping probabilities can be taken symmetric ($w_{i,i+1}=w_{i+1,i})$ or non
symmetric, and will be chosen according to the following typical probability
distribution:
\begin{equation}
\rho(w)=2^{1-\beta}(1-\beta)w^{-\beta}\theta(w)\theta(1/2-w),
\label{rho}
\end{equation}
where $\beta<1$ measures the quenched disorder strength.

The case $\gamma=1$ (no trapping) has been extensively studied
\cite{alex,bouchaud,sinai}. In the symmetric case \cite{alex,bouchaud}, one
observes  anomalous diffusion, $\langle x^2(t)\rangle\sim t^{2\nu}$, with
$\nu$ depending continuously on $\beta$. The  return probability
$P_s(t)=\langle p_{i,i}(t)\rangle$, that is the probability of being at site
$i$ at time $t$ having started at site $i$, decays as $P_s(t)\sim t^{-d_s/2}$,
where $d_s$ is the spectral dimension \cite{alex,bouchaud}. The Sinai model
\cite{sinai} describes the generic non symmetric case and displays
logarithmically slow diffusion and other peculiar properties.

In the presence of trapping ($0\leq\gamma<1$), the problem has been studied
essentially by numerical means \cite{gia1,gia2}. In addition to $P_s(t)$, one
can define the normalized return probability $P(t)$ as,
\begin{equation}
P(t)=\left\langle\frac{p_{i,i}(t)}{\sum_jp_{i,j}(t)}\right\rangle.
\end{equation} 
Note that in order to keep the notations simple, it is understood that 
$\langle\ldots\rangle$ involves an average over the disorder $and$ the
considered site $i$.
The proper way of defining $\langle x^2(t)\rangle$ is now,
\begin{equation}
\langle x^2(t)\rangle=\left\langle
\frac{\sum_j p_{i,j}(t)(j-i)^2}{\sum_jp_{i,j}(t)}\right\rangle,
\label{diffu}
\end{equation}
only taking into account surviving particles. With these new definitions,
$\langle x^2(t)\rangle\sim t^{2\nu}$, with $2\nu\approx 1.25$, seemingly
independent of $\gamma$ and the disorder strength $\beta$ \cite{gia2}. $P(t)$
decays as a power-law, $P(t) \sim t^{-\alpha}$, with $\alpha\approx 0.59$,
also independent of $\gamma$ and $\beta$. Due to trapping, the survival return
probability $P_s(t)=\langle p_{i,i}(t)\rangle$ decays much faster, and the
authors of \cite{gia2} gave an heuristic argument leading to,
\begin{equation}
\log P_s(t)\sim -\sqrt{t},
\label{ps}
\end{equation}
in qualitative agreement with numerical simulations. Moreover the probability
distributions  of quantities such as $p_{i,j}(t)$ have been shown
numerically to be very broad, leading to non self-averaging effects.

Some of the peculiar properties of this model have been related to the possible
existence of multifractal distributions for quantities such as the escape
probability (see below) \cite{gia1}, in analogy \cite{murth} to what has been
observed for the Sinai model \cite{sinai}.

For instance, let us consider the probability $G_{i,i+1}(t)$ that a particle
makes a first passage from a site $i$ to a site $i+1$ in $t$ steps. This obeys
the following master equation \cite{gia1}:
\begin{equation}
G_{i,i+1}(t)=w_{i,i+1}\delta_{t,1}+w_{i,i-1}G_{i-1,i+1}(t)
+\gamma(1-w_{i,i+1}-w_{i,i-1})G_{i,i+1}(t-1),
\label{master}
\end{equation} 
with the boundary conditions, 
\begin{equation}
G_{0,1}(t)=w_{0,1}\delta_{t,1}
+\gamma(1-w_{0,1})G_{0,1}(t-1),
\end{equation}
and $G_{i,i+1}(t)=0$, for $i<0$. Now the escape probability is defined as,
\begin{equation}
x_i=\sum_{t=0}^{+\infty}G_{i,i+1}(t),
\label{xi}
\end{equation}
and the authors of \cite{gia1} claim to have found a multifractal distribution
for this quantity, after averaging over $i$ and the quenched disorder.

In the next section, we show analytically that this escape probability has  in
fact a well defined distribution and that the apparent multifractality can be
ascribed to peculiar features of this distribution. In addition, we use some
of these results in section {\bf IV} to derive the exact large time
behavior of $P_s(t)$, by computing the properties of the
survival probability distribution. We also give an exact bound on the density
of particles, which fully confirms our result:
\begin{equation}
\log P_s(t)\sim -t^{1/3}\log^{2/3}(t).
\label{ps1}
\end{equation}

Sections {\bf III.C} and {\bf IV.B}, where a rigorous method to analyse
Lifshitz-like tails is introduced, are rather technical. The reader more
interested in the physical consequences of these results could skip these
technicalities and take the qualitative arguments given in the beginning of
these sections for granted.

\section{Escape probability distribution: non symmetric case}

\subsection{Preliminaries}

Let us show that $x_{i+1}$ can be simply evaluated from the knowledge of
$x_i$, a calculation already appearing in \cite{murth2,gia1}.
We define the generating functions, 
\begin{equation}
\hat G_{i,j}(z)=\sum_{t=0}^{+\infty}z^tG_{i,j}(t),
\end{equation}
with $x_i=\hat G_{i,i+1}(z=1)$. The convolution theorem ensures that, 
\begin{equation}
\hat G_{i-1,i+1}(z)=\hat G_{i-1,i}(z)\hat G_{i,i+1}(z).
\end{equation}
Using the master equation \eqn{(\ref{master})} this straightforwardly leads to
the iterated map:
\begin{equation}
x_i=\frac{p_i}{1-\gamma(1-p_i-q_i)-q_ix_{i-1}}, \quad i\geq 1,
\label{map}
\end{equation}
and,
\begin{equation}
x_0=\frac{p_0}{1-\gamma(1-p_0)}, \quad i\geq 1,
\end{equation}
where $p_i=w_{i,i+1}$ and $q_i=w_{i,i-1}$ are random variable of identical
distribution $\rho$ given in \eqn{(\ref{rho})}. Note that these variables are
independent only in the non symmetric model, as the extra constraint
$q_i=p_{i-1}$ holds in the symmetric case.

To the price of lengthier calculations, we have checked that the asymptotic
behaviors of the observables of interest are not  affected by the value of
$\gamma$, provided trapping is not suppressed ($\gamma<1$). This is in
agreement with the numerical simulations performed in \cite{gia1,gia2}. From
now on, we therefore restrict our analytic study to $\gamma=0$.

In the case $\gamma=0$, the maximum possible value for $x$ should satisfy,
\begin{equation}
x_{\rm max}=\frac{p_{\rm max}}{1-q_{\rm max}x_{\rm max}},
\end{equation}
with $p_{\rm max}=q_{\rm max}=1/2$, leading to $x_{\rm max}=1$. 

\subsection{Equation for the stationary distribution: non symmetric case}

As noticed above, the $p_i$'s and $q_i$'s are independent variable in the non
symmetric case.
Let us now assume that for large $i$ the probability distribution of $x_i$
exists and becomes independent of $i$, in contradiction with the numerical
claim of \cite{gia1}. The stationarity of the distribution is exploited by
expressing that the distribution of $x_i$ should be the same as that of
$x_{i-1}$ (at least for large $i$), leading to:
\begin{equation}
f(x)=\int_0^1dy \int_0^{1/2}dp \int_0^{1/2}dq
f(y)\rho(p)\rho(q)\delta\left(x-\frac{p}{1-qy}\right).
\end{equation}
This equation, though apparently very complicated, can be in fact exploited
quite precisely. First consider $x\leq 1/2$, $p=x(1-qy)$ is then always within
the integration domain $[0,1/2]$, for any $q\in [0,1/2]$ and any $y\in
[0,1]$. We then find,
\begin{eqnarray}
f(x)&=\int_0^1dy  \int_0^{1/2}dq f(y)\rho(x(1-qy))\rho(q)&(1-qy),\\
    &=2^{1-\beta}(1-\beta)x^{-\beta}
    \int_0^1dy  \int_0^{1/2}dq f(y)&(1-qy)^{1-\beta}\rho(q),
    \label{xinf}
\end{eqnarray}
that is a pure power-law behavior.

For $x>1/2$, imposing $p=x(1-qy)\in [0,1/2]$ leads to new integration bounds:
\begin{eqnarray}
f(x)&=2^{2(1-\beta)}(1-\beta)^2x^{-\beta}\int_{2-x^{-1}}^1dy 
\int_{(1-(2x)^{-1})y^{-1}}^{1/2}dq f(y)(1-qy)^{1-\beta}q^{-\beta},\\
    &=2^{2(1-\beta)}(1-\beta)^2x^{-\beta}\int_{2-x^{-1}}^1dy 
\int_{1-(2x)^{-1}}^{y/2}dq f(y)y^{\beta-1}(1-q)^{1-\beta}q^{-\beta}.
\label{xsup}
\end{eqnarray}
Introducing $g(x)=x^\beta f(x)$, one can differentiate twice this equation
with respect to the variable $x$, leading to the following non local
differential equation for $g$, valid for $x\in [1/2,1]$:
\begin{eqnarray}
g^{\prime\prime}(x)+\left[\frac{3-\beta}{x}+\frac{\beta}
{x^2\left(2-x^{-1}\right)}\right]
g^{\prime}(x)=(1-\beta)^2x^{\beta-5}\left[2-x^{-1}\right]^{-(1+\beta)}
g\left(2-x^{-1}\right).
\label{diff}
\end{eqnarray}
The structure of this equation is quite unfamiliar as the LHS linear
differential operator is determined by the value of $g$ at $x'=2-x^{-1}$ in
the RHS.

$g(x)=x^\beta f(x)$ being zero for $x'<0$ (that is $x\in [0,1/2[$), this
equation actually reproduces that $g'(x)$ should be zero for $x\in [0,1/2[$,
in agreement with \eqn{(\ref{xinf})}. Then, the knowledge of $g(x')$ for
$x'\in [0,1/2[$ determinates the LHS on the interval $x\in [1/2,2/3]$. $g$ is
then determined by imposing that it is continuous at $x=1/2$, and a
consistency equation at $x=2/3$ (see below). This procedure can be iterated.
Consider $u_0=0$, and $u_{n+1}=1/(2-u_n)$ (that is $u_n=2-u_{n+1}^{-1}$), the
knowledge of $g$ on the interval $[u_{n-1},u_n[$ and the continuity condition
at $x=u_n$ for $g$ and its first derivative fully determine the function $g$
on the next interval $[u_{n},u_{n+1}[$. $u_n$ can be exactly computed by
induction leading to,
\begin{equation}
\varepsilon_n=1-u_n=\frac{1}{n+1}.
\end{equation}
As can be expected, $u_n\to x_{\rm max}=1$, when $n\to +\infty$. 

This recursion process and the form of \eqn{(\ref{diff})} ensures that $g$ is
infinitely differentiable on the interval $]u_n,u_{n+1}[$. Moreover, if $g$ is
$d_{n+1}$ times differentiable at $x=u_{n+1}$, it is at least $d_{n+1}+1$
times differentiable at $x=u_{n+2}$. This shows that the continuity condition
for $g$ and its first derivative suffices to determine $g$ on the interval
$[u_n,u_{n+1}[$ for $n>1$. However, the knowledge of $g$ on the interval
$[0,1/2[$ is not sufficient to determine $g$ on the next interval $[1/2,2/3[$,
as the derivative of $g$ is not continuous at $x=1/2$. For instance, for
$\beta>0$, one has $g'\left(\frac{1}{2}^-\right)=0$, whereas it can be shown
that $g'(1/2+\varepsilon)\sim \varepsilon^{-\beta}\log(\varepsilon)$. As the
differential equation in \eqn{(\ref{diff})} is of the second order type, one
of the integration constants remains unknown, the other being determined by the
continuity condition at $x=1/2$. We are thus left with a classical shooting
problem, where $g'(2/3)$ will be fixed by asking that the distribution
vanishes at $x=1$.

Let us  make this point clearer in the case $\beta=0$, for which the first
iteration can be explicitly performed starting from $g(x)=a=g(0)$ for $x\in
[0,1/2[$, where the constant $a$ is given in \eqn{(\ref{xinf})}. This constant
can be eventually calculated once the full distribution is known up to this
overall constant, as it will ensure the proper normalization of the
distribution $f(x)=x^{-\beta}g(x)$. For $x\in [1/2,2/3]$, we get,
\begin{equation}
\frac{g(x)}{g(0)}=\frac{1}{4x^2}+\frac{1}{x}-2-2\left(1-\frac{1}{4x^2}\right)
\left(\log\left[\frac{x}{2(2x-1)}\right]+c\right),
\end{equation}
where $c=-\frac{8}{27}\cdot\frac{g'(2/3)}{g(0)}$. Note that in this case, we
indeed find that $g'(1/2+\varepsilon)\sim\log(\varepsilon)$, leading to an
infinite derivative at $x=(1/2)^+$. This form for $g$, valid on the interval
$[1/2,2/3]$, the result of \eqn{(\ref{diff})}  and the continuity condition
for $g$ and $g'$ at $u_n$ (for $n>1$) leads to the full determination of $g$.
Then, a  proper choice of the constant $c$ ensures that $f(1)=0$.

In fig. 1, $g(x)$ (equal to $f(x)$ for $\beta=0$) is shown for different
values of $c$. For the optimal choice for $c$, it coincides perfectly with the
distribution obtained by directly iterating \eqn{(\ref{map})}.

In fig. 2, $g(x)=x^\beta f(x)$ is shown for $\beta=-1/2$ (weak disorder) and 
$\beta=1/2$ (strong disorder). The small $x$ behavior for $f$ is
confirmed, and the distributions are again in perfect agreement with the
numerical integration of \eqn{(\ref{diff})}.

\subsection{Lifshitz tail at the edge of the spectrum}

Note that all these distributions seem to vanish well before $x_{\rm max}=1$.
We will show below that this is not the case and that this apparent behavior
can be accounted by the fact that,
\begin{equation}
f(1-\varepsilon)\sim \varepsilon^{-4/\varepsilon},\qquad {\rm when}\quad
\varepsilon\to 0.
\label{tail}
\end{equation}
We shall see in section {\bf III.E} that the apparent multifractal
properties of the $x_i$'s observed in \cite{gia1} are partly due to this
phenomenon.

Let us give a qualitative justification of \eqn{(\ref{tail})}. Taking
$x=1-\varepsilon$, and expecting a very fast decay of $g$ at $x=1$, the two
most singular terms in \eqn{(\ref{diff})} should be $g^{\prime\prime}$ and the
RHS. It can then be checked that the $ansatz$ $g(1-\varepsilon)\sim
\varepsilon^{-\alpha/\varepsilon}$ is solution of \eqn{(\ref{diff})} (up to
subleading multiplicative logarithmic terms) if one takes $\alpha=4$.

Still, actually solving \eqn{(\ref{diff})}, even in the limit $x\to 1$,
remains a formidable task and the previous argument should be taken with care.
However, we can justify rigorously  the fast decaying tail of $f$ at $x=1$,
finding a result fully compatible with \eqn{(\ref{tail})}. 

Consider $P(\varepsilon)=\int_{1-\varepsilon}^1f(x)\,dx$, the probability to
have $x_i>1-\varepsilon$. In a way similar to that leading to
\eqn{(\ref{xsup})}, we find,
\begin{equation}
P(\varepsilon)=\int_{1-h(\varepsilon)}^1 dy\int_{(1-h(\varepsilon))/2y}^{1/2}dq
\int_{(1-\varepsilon)(1-qy)}^{1/2}dp f(y)\rho(p)\rho(q), 
\end{equation}
where $h(\varepsilon)=\varepsilon/(1-\varepsilon)$. For small $\varepsilon$,
we expect $P(\varepsilon)$ to be very small as $p$ and $q$ are to be taken
close to $1/2$ and, simultaneously, $y$  must be close to 1 (see
\eqn{(\ref{map})}). In the vicinity of $(p,q)=1/2$, the distribution $\rho$ is
smooth and roughly constant (see \eqn{(\ref{rho})}). We thus get:
\begin{eqnarray}
P(\varepsilon)&\sim 4(1-\beta &)^2\int_{1-h(\varepsilon)}^{1} dy 
\int_{(1-h(\varepsilon))/2y}^{1/2}dq 
\int_{(1-\varepsilon)(1-qy)}^{1/2}dp f(y),\\
&\sim \frac{(1-\beta)^2}{2}&\int_{1-h(\varepsilon)}^1 dyf(y)\left[1-y-
h(\varepsilon)\right]^2.
\label{ineg} 
\end{eqnarray}
From now, we use the symbol $\sim$ in its true mathematical sense, such that
$a(\varepsilon)\sim b(\varepsilon)$ means that
$a(\varepsilon)/b(\varepsilon)\to 1$, when $\varepsilon\to 0$.
For sufficiently small $\varepsilon$, this leads to the following upper bound:
\begin{equation}
P(\varepsilon)\leq a_+\int_{1-h(\varepsilon)}^1 dyf(y)\left[1-y-
h(\varepsilon)\right]^2,
\end{equation}
valid for any constant $a_+>\frac{(1-\beta)^2}{2}$. We then get,
\begin{equation}
P(\varepsilon)\leq a_+\int_{1-h(\varepsilon)}^{1-\varepsilon}
dyf(y)\left[\varepsilon- h(\varepsilon)\right]^2+a_+\int_{1-\varepsilon}^1
dyf(y)h(\varepsilon)^2,
\end{equation}
where we have used the fact that, 
\begin{eqnarray}
&\left[1-y-h(\varepsilon)\right]^2\leq &\left[\varepsilon-
h(\varepsilon)\right]^2\sim\varepsilon^4, \quad {\rm for}\quad
y\in [1-h(\varepsilon),1-\varepsilon],\\
&\left[1-y-h(\varepsilon)\right]^2\leq &
h(\varepsilon)^2\sim\varepsilon^2, \quad {\rm for}\quad
y\in [1-\varepsilon,1].
\end{eqnarray}
Thus, there exists two constants $a_1$ and $a_2$, such that,
\begin{equation}
P(\varepsilon)\leq a_1\varepsilon^4P(h(\varepsilon))
+a_2\varepsilon^2P(\varepsilon).
\end{equation}
Finally, this last inequality shows that for $\varepsilon$ sufficiently small,
there exists a constant $A_+>0$ such that,
\begin{equation}
P(\varepsilon)\leq A_+\varepsilon^4P(h(\varepsilon)).
\label{req1}
\end{equation}
On the other hand, using again \eqn{(\ref{ineg})} and
choosing a sufficiently small $\delta$ to be determined below, we have,
\begin{eqnarray}
P(\varepsilon)&\geq \frac{(1-\beta)^2}{2}
\int_{1-h(\varepsilon)(1+\delta)}^1 dyf(y)&\left[1-y-
h(\varepsilon)\right]^2, \\
&\geq A_- P(h(\varepsilon)(1+\delta))\varepsilon^2\delta^2,
\label{req2}
\end{eqnarray}
again valid for any constant $A_-<\frac{(1-\beta)^2}{2}$, for small enough 
$\varepsilon$. In the following, we take $\delta=\varepsilon^\alpha$, and will
fix the constraint on $\alpha$ later.

Let us now start from a small enough $\varepsilon$ such that the inequalities
in \eqn{(\ref{req1})} and \eqn{(\ref{req2})} hold. Taking
$\varepsilon_0^+=\varepsilon_0^-=\varepsilon$, we then define
$\varepsilon_n^+$ and $\varepsilon_n^-$ by the recursion relations,
\begin{equation}
\varepsilon_n^+=h(\varepsilon_{n+1}^+)=
\frac{\varepsilon_{n+1}^+}{1-\varepsilon_{n+1}^+},\qquad
\varepsilon_n^-=\varepsilon_{n+1}^-\frac{1+({\varepsilon_{n+1}^-})^\alpha}
{1-\varepsilon_{n+1}^-}.
\end{equation}
Both sequences go to $\varepsilon=0$. These recursion relations can also be
rewritten,
\begin{eqnarray}
&\frac{1}{\varepsilon_{n+1}^+}-\frac{1}{\varepsilon_{n}^+}=&1,\label{equ1}\\
&\frac{1}{\varepsilon_{n+1}^-}-\frac{1}{\varepsilon_{n}^-}=&1+
O\left({\varepsilon_{n+1}^-}^{(\alpha-1)}\right).
\label{equ}
\end{eqnarray}
Thus, for any $\alpha>1$, both sequences are equivalent to $n^{-1}$
(by applying Cesaro's mean theorem):
\begin{equation}
\varepsilon_{n}^+\sim \varepsilon_{n}^- \sim \frac{1}{n}.
\end{equation}
Then, by iterating the recursion relations of \eqn{(\ref{req1},\ref{req2})},
we get:
\begin{eqnarray}
\log(P(\varepsilon_{n}^+))\leq
4\sum_{k=0}^{n-1}\log(\varepsilon_{n}^+)+n\log(A_+)+\log(P(\varepsilon))
\sim -4n\log(n),\\
\log(P(\varepsilon_{n}^-))\geq
2(1+\alpha)\sum_{k=0}^{n-1}\log(\varepsilon_{n}^-)
+n\log(A_-)+\log(P(\varepsilon))
\sim -2(1+\alpha)n\log(n).
\end{eqnarray}
Finally, using \eqn{(\ref{equ1},\ref{equ})} and the fact that $P(\varepsilon)$
is a continuous and increasing function of $\varepsilon$, and as $\alpha$ can
be arbitrary close to 1, we get that for any arbitrary small $\eta>0$, there
exists $\hat\varepsilon>0$, such that for any $0<\varepsilon<\hat\varepsilon$,
\begin{equation}
-\frac{4(1-\eta)}{\varepsilon}\log(\varepsilon)\leq 
-\log(P(\varepsilon))\leq -\frac{4(1+\eta)}{\varepsilon}\log(\varepsilon),
\end{equation}
which leads to our final result:
\begin{equation}
-\log(P(\varepsilon))\sim-\log(f(1-\varepsilon))\sim
-\frac{4}{\varepsilon}\log(\varepsilon), 
\quad{\rm when}\quad \varepsilon\to 0,
\label{eqth}
\end{equation}
which is a more precise and rigorous statement than that of
\eqn{(\ref{tail})}. It can also be shown that the subleading terms in
\eqn{(\ref{eqth})} are {\it a priori} of order
$\log(\log(1/\varepsilon))/\varepsilon$. In practice, these strong subleading
corrections and the very fast decay of the distribution near $x=1$ make the
quantitative numerical confirmation of \eqn{(\ref{eqth})} quite difficult.

\subsection{Escape probability distribution: symmetric case}

In this section we are interested in the symmetric version of our model for
which $q_i=p_{i-1}$. The approach is slightly different from that of the
previous case but is definitively in the same spirit. As a consequence, less
attention will be attached to rigorous arguments, although they can be adapted
without any difficulty to this problem.

The map now becomes,
\begin{equation}  
x_i=\frac{p_i}{1-p_{i-1}x_{i-1}},
\end{equation}
which shows that the novel variable $y_i=p_ix_i$ satisfies the recursion,
\begin{equation}  
y_i=\frac{p_i^2}{1-y_{i-1}}.
\label{mapy}
\end{equation} 
$u=p^2$ has a distribution $\sigma(u)$ satisfying,
\begin{equation}
\sigma(u)=\int_{0}^{1/2}dw\,\rho(w)\delta(u-w^2)=2^{-\beta}(1-\beta)
u^{-(1+\beta)/2}\theta(u)\theta(1/4-u).
\label{sig}
\end{equation}
Using \eqn{(\ref{mapy})}, we find that $y$ is always in $[0,1/2]$, and that
for $y\in [0,1/4]$, the probability distribution $F(y)$ of $y$ satisfies,
\begin{equation}
F(y)=2^{-\beta}(1-\beta)y^{-(1+\beta)/2}\int_0^{1/2}F(y')(1-y')^{(1-\beta)/2}
\,dy',
\end{equation}
that is a pure power law behavior. For $y>1/4$, and proceeding along the same
line as in the non symmetric case, we find that $F(y)$
satisfies the following non local differential equation:
\begin{equation}
F'(y)+\frac{1+\beta}{2y}F(y)=\frac{1-\beta}{8y^3}F\left(1-\frac{1}{4y}\right).
\end{equation}
Again, the knowledge of $F(y)$ on the interval $[0,1/4]$, permits the
determination of the distribution on the next interval $[1/4,1/3]$, and by
recursion on each of the intervals of the form $[u_n,u_{n+1}]$, with
$u_n=\frac{n}{2(n+1)}$.

Now, let us analyze the behavior of $P(\varepsilon)=
\int_{1/2-\varepsilon}^{1/2} F(y)\, dy$, for small $\varepsilon$. Defining
$h(\varepsilon)=\varepsilon/(1-2\varepsilon)$, we easily find that there
exists two constants $c$ and $C$ (which can be actually determined) such that,
\begin{equation}
P(\varepsilon)\sim c\int_0^{h(\varepsilon)}(h(\varepsilon)-z)F(1/2-z)\,dz\sim
C\varepsilon^2P(h(\varepsilon)).
\end{equation}
The last estimate is obtained using the same type of inequalities as in the
non symmetric case. Again defining,
$\varepsilon_{n+1}= \varepsilon_n/(1+2\varepsilon_n)$, and using the fact that
$\varepsilon_n\sim {(2n)^{-1}}$, we find that,
\begin{equation}
\log(P(\varepsilon_n))\sim 2\sum_{k=1}^n \log(\varepsilon_k)\sim
\frac{\log(\varepsilon_n)}{\varepsilon_n}, 
\end{equation}
which finally shows that,
\begin{equation}
\log(P(\varepsilon))\sim\log(F(1/2-\varepsilon))\sim 
\frac{\log(\varepsilon)}{\varepsilon}.
\label{eqforF}
\end{equation} 
From the knowledge of the properties of $F(y)$, we can access to that of
$f(x)$, the stationary distribution of the $x_i$'s, using the relation,
\begin{equation}
f(x)=\int_0^{1/2}dp\int_0^{1/2}dy
\rho(p)F(y)\delta\left(x-\frac{p}{1-y}\right).
\label{convol}
\end{equation} 
Let us exhibit the main properties of $f$ which will be completely similar as
what was obtained in the non symmetric version of the model.
For $x\in [0,1/2]$, \eqn{(\ref{convol})} leads to,
\begin{equation}
f(x)=2^{1-\beta}(1-\beta)x^{-\beta}\int_0^{1/2} F(y)(1-y)^{1-\beta}\, dy,
\end{equation}
which shows that $f(x)$ is simply proportional to $x^{-\beta}$ as in the non
symmetric case.  For $x\in [1/2,1]$, we obtain,
\begin{equation}
g'(x)=(x^\beta f(x))'=\frac{1-\beta}{2x^{3-\beta}}F\left(1-\frac{1}{2x}\right).
\end{equation}
Using \eqn{(\ref{eqforF})}, we finally conclude that,
\begin{equation}
\log(f(1-\varepsilon))\sim \log\left(F
\left(1-\frac{1}{2(1-\varepsilon)}\right)\right)\sim \frac{2}{\varepsilon}
\log(\varepsilon),
\end{equation}
that is a similar behavior as was found in the non symmetric case (see
\eqn{(\ref{eqth})}) up to the factor 4 which is replaced by a factor 2. The
physical interpretation of this smaller coefficient is quite clear: in the non
symmetric case, for $x_i$ to be close to 1, one must have $x_{i-1}$, $p_i$ and
$q_i$ close to their maximal value. In the symmetric case,  for $x_i$ to be
close to 1, only $x_{i-1}$ and  $p_i$ must be close to their maximal value, as
$q_i=p_{i-1}$ is automatically close to $1/2$ as $x_{i-1}$ is close to 1. This
extra constraint explains why $f(x)$ decays faster in the non symmetric case
than in the symmetric case which is confirmed numerically on fig. 3.

Note  that the non symmetric case could have been treated by the same
technics as in this subsection by replacing  the distribution $\sigma$ of 
$p_i^2$ by the distribution ${\hat\sigma}$ of $p_iq_{i+1}$. 

Finally, we can conclude that up to a few irrelevant details, the symmetric
and non symmetric models seem to share exactly the same properties. This is
apparently surprising as the corresponding models without trapping are
drastically different \cite{bouchaud}. This intriguing property is confirmed
and explained physically in section {\bf IV.D}.

\subsection{Explanation of the apparent multifractality}

In the preceding subsections, we have obtained a puzzling numerical result
(see fig. 1-3):
although we have shown that the maximum value $x_{\rm max}=1$ must be
attained, the numerical maximum value effectively obtained after $2\times
10^9$ iterations of the map \eqn{(\ref{map})} is typically of order $x_{\rm
max,eff}\approx 0.73\sim 78$, for the three values of $\beta$ actually tested
in the non symmetric case. This
apparent paradox can be explained by the sharp decay of the distribution
$f(x)$. $x_{\rm max,eff}$ can be estimated by considering that after $N$
iterations of the map, 
\begin{equation}
\int_{x_{\rm max,eff}}^1f(x)\,dx\sim N^{-1},
\end{equation}
which is the smallest non zero value that this integral can take
(the integrated distribution increasing by elementary steps of height $N^{-1}$).
Using \eqn{(\ref{eqth})}, we find that $\varepsilon=1-x_{\rm max,eff}$ must
satisfy,
\begin{equation}
-\frac{\log(\varepsilon)}{\varepsilon}\sim\frac{1}{4}\log(N).
\end{equation}
If we now take $N=2\times 10^9$, the above estimate gives $x_{\rm
max,eff}\approx 0.75$, in fair agreement with the observed range
of numerical effective values for  $x_{\rm max}$.

In the limit of very large $N$, \eqn{(\ref{eqth})} also leads to 
the leading order estimate,
\begin{equation}
1-x_{\rm max,eff}\approx\frac{4\log(\log(N))}{\log(N)},
\end{equation}
which goes to zero very slowly. 

In \cite{gia1}, the fact that $x_{\rm max}=1$ was not recognized. The
authors actually computed the multifractal distribution of the $x_i$'s in the
interval $[0,x_{\rm max,eff}]$. This interval was cut into $n$ equal length
intervals, and $\rho_j$ was defined as the fraction of the total number of the
$x_i$'s which belongs to the $j^{\rm th}$ interval. One then defines
\cite{fract},
\begin{equation}
Z(q,n)=\sum_{i=1}^n\rho_i^q\sim n^{-\tau(q)},
\end{equation}
the last equivalent defining the scaling exponent $\tau(q)$ associated to the
$q^{\rm th}$ moment. 
Let us first derive the exact expression of $\tau(q)$ for a given choice of
$\beta>0$ (strong disorder), and taking the actual value of $x_{\rm max}=1$.
For $0\leq q<1/\beta$, the function $f(x)^q$ is integrable, so that,
\begin{equation}
Z(q,n)=n^{-(q-1)}\times\frac{1}{n}\sum_{i=1}^n(n\rho_i)^q\sim n^{-(q-1)}
\int_0^1f(x)^q\,dx.
\end{equation}
This shows that in this regime, $\tau(q)=q-1$. For $q>1/\beta$,
the function $f(x)^q$ is no longer integrable due to the power-law divergence
at $x=0$:
\begin{equation}
Z(q,n)=n^{-(q-1)}\times\frac{1}{n}\sum_{i=1}^n(n\rho_i)^q\sim n^{-(q-1)}
\int_{1/n}^1f(x)^q\,dx\sim n^{-(1-\beta)q}.
\end{equation}
In this regime, we thus find $\tau(q)=(1-\beta)q$. Morover, for $q=1/\beta$,
we find,
\begin{equation}
Z(q=1/\beta,n)\sim n^{-(1/\beta-1)}\int_{1/n}^1f(x)^{1/\beta}\,dx
\sim n^{-(1/\beta-1)}\log(n),
\end{equation}
which shows that in the region $q\approx 1/\beta$, the numerical determination
of $\tau(q)$ will be strongly affected by a logarithmic slow cross-over.
Finally, strictly speaking, $\tau(q)$ is not defined for negative $q$, due to
the essential singularity at $x_{max}=1$. But if one computes the multifractal
scaling exponents by restraining the study on the interval $[0,x_{\rm
max,eff}]$ (as in \cite{gia1}), we then recover $\tau(q)=q-1$, for negative
values of $q$ as well.

The two linear regimes for $\tau(q)$ are clearly visible on the fig. 4 of
\cite{gia1}, with the predicted slopes and transition point $q=1/\beta$. Of
course, numerically, the change in slope (from $\tau'(q)=1$ to
$\tau'(q)=1-\beta$) is found to be smooth, partly due to the logarithmic
correction around $q=1/\beta$. Then, the spectrum of singularities defined as
the Legendre transform of $\tau(q)$ \cite{fract},
\begin{equation}
s(\alpha)=\alpha q-\tau(q),\quad \alpha=\tau'(q)
\end{equation}
apparently yields a non trivial spectrum, whereas the actual one is
concentrated on two points: $\alpha=1$, with a support of fractal dimension
$s(1)=1$, reflecting that for almost all values of $x$ the distribution $f(x)$
is actually continuous, and $\alpha=1-\beta$, with a support of fractal
dimension $s(1-\beta)=0$, which simply results from the $x=0$ singularity of
the distribution $f(x)$.

The moral that we can draw from this is that one must be very careful when
dealing with multifractal analysis, especially if there is no special
reason to expect a multifractal spectrum. Similar problems were encountered
by the author of \cite{nagatini}, who obtained an apparent multifractal
spectrum in a model which can actually be solved exactly \cite{takayasu}, and
for which it can be shown that the true multifractal spectrum is of the same
type as above. Similar doubts can be raised on the findings of multifractal
spectra in certain biological systems or in the field of finance \cite{bou2},
where simple power-law distributions can lead to such apparent behaviors.

\section{Large time behavior of the survival return probability}

\subsection{General results}

Following the tracks of \cite{gia2}, let us evaluate the survival return, or
more exactly, its discrete time Laplace transform. We consider the symmetric
model but the non symmetric case can be treated in the same spirit, leading to
exactly the same results.

Thus consider,
\begin{equation} 
\hat p_{0,i}(\omega)=\sum_{t=0}^{+\infty}\frac{p_{0,i}(t)}{(1+\omega)^{t+1}}.
\end{equation}
It can be shown \cite{gia2} that $\hat p_{0,i}(\omega)$ satisfies the equation
(see \eqn{(\ref{master})}),
\begin{equation} 
\hat p_{0,i}(\omega)=w_{i-1,i}\hat p_{0,i-1}(\omega)+
w_{i,i+1}\hat p_{0,i+1}(\omega)+(1+\omega)^{-1}\delta_{i,0}.
\end{equation}
Then, the variables $\phi_i^+(\omega)$ and $\phi_i^-(\omega)$, defined
respectively  for $i>0$ and $i<0$ by,
\begin{equation} 
\phi_i^+(\omega)=\frac{w_{i-1,i}}{1+\omega}\cdot
\frac{\hat p_{0,i}(\omega)}{\hat p_{0,i-1}(\omega)},
\quad
\phi_i^-(\omega)=\frac{w_{i,i+1}}{1+\omega}\cdot
\frac{\hat p_{0,i}(\omega)}{\hat p_{0,i-1}(\omega)},
\end{equation}
satisfy the same recursion, reminiscent of that of \eqn{(\ref{map})}. For
instance,
\begin{equation} 
\phi_i^+(\omega)=\frac{\mu_{i-1}^2}{1-\phi_{i+1}^+(\omega)},\quad
{\rm with}\quad \mu_{i-1}=\frac{w_{i-1,i}}{1+\omega},
\label{map2}
\end{equation}
with a similar equation for $i<0$. It can then be shown that,
\begin{equation}
\langle p_{0,0}(\omega)\rangle=\int_{0}^{\phi_{\rm max}(\omega)}d\phi^+
\int_{0}^{\phi_{\rm max}(\omega)}d\phi^-\,\Pi_\omega(\phi^+)\Pi_\omega(\phi^-)
\frac{\theta(1-\phi^+-\phi^-)}{1-\phi^+-\phi^-},
\label{survi}
\end{equation}
where $\Pi_\omega(\phi)$, is the expected stationary distribution of $\phi^+$
and $\phi^-$.

From now, we follow the method of section {\bf III} to evaluate the
behavior of the distribution  $\Pi_\omega(\phi)$ close to $\phi=\phi_{\rm
max}$, and for small $\omega$ (as we are mainly interested in the large time
behavior of $P_s(t)$). From \eqn{(\ref{map2})}, $\phi_{\rm max}(\omega)$ can
be easily calculated:
\begin{equation}
\phi_{\rm max}(\omega)=\frac{1}{2}\left(1-\sqrt{1-(1+\omega)^{-2}}\right)
=\frac{1}{2}-\sqrt{\frac{\omega}{2}}+O(\omega^{3/2}).
\end{equation}
Moreover, using again \eqn{(\ref{map2})}, we find that $\Pi_\omega(\phi)$,
satisfies the following self-consistent equation:
\begin{equation}
\Pi_\omega(\phi)=\int_0^{1/4}dx \int_0^{\phi_{\rm max}(\omega)}d\phi'\,
\sigma(x)\Pi_\omega(\phi')\delta\left(\phi-\frac{x}{(1+\omega)^2(1-\phi')}
\right), 
\label{pi}
\end{equation}
where the distribution $\sigma(x)$ of $x=\mu^2(1+\omega)^2$ is given by
\eqn{(\ref{sig})}.

\eqn{(\ref{map2})} can be first solved for $\phi<\frac{1}{4(1+\omega)^2}$,
leading to a pure power-law behavior (as encountered in section {\bf III}):
\begin{equation}
\Pi_\omega(\phi)=2^{-\beta}(1-\beta)(1+\omega)^{-(1+\beta)}
\phi^{-(1+\beta)/2}\int_0^{\phi_{\rm max}(\omega)} d\phi'\,
\frac{\Pi_\omega(\phi)}{(1-\phi')^{(1+\beta)/2}}.
\end{equation}
For $\phi>\frac{1}{4(1+\omega)^2}$, one can differentiate \eqn{(\ref{pi})}
(noticing that $\phi'\in \left[1-\frac{1}{4(1+\omega)^2\phi}\right]$), which
leads to the following differential equation for $\Pi_\omega(\phi)$:
\begin{equation}
\left[\phi^{(1+\beta)/2}\Pi_\omega(\phi)\right]'=-\frac{1-\beta}{2}
(1+\omega)^{-(5+\beta)/2}\phi^{-(3-\beta)/2}
\Pi_\omega\left(1-\frac{1}{4(1+\omega)^2\phi}\right). 
\label{eqdif2}
\end{equation}
The leading asymptotics of $\Pi_\omega(\phi)$ close to  $\phi=\phi_{\rm
max}(\omega)$ can be calculated rigorously adapting the method of section {\bf
III.C}. Here, we first derive this result by a less rigorous method already
mentioned in section {\bf III.A}, consisting in keeping only the most singular
terms in the differential equation \eqn{(\ref{eqdif2})}, leading to:
\begin{equation}
\Pi_\omega^{\prime}(\phi)\sim A 
\Pi_\omega\left(1-\frac{1}{4(1+\omega)^2\phi}\right)\sim A 
\Pi_\omega\left(\phi_{\rm max}(\omega)-\frac{1-\phi_{\rm max}(\omega)}
{\phi_{\rm max}(\omega)}\varepsilon\right),
\label{anz2}
\end{equation}
where $A$ is a computable positive constant, and the explicit equation for 
$\phi_{\rm max}(\omega)$ was used. One can try an $ansatz$ form for
$\Pi_\omega(\phi)$ which satisfies this equation up to multiplicative
logarithmic terms. We find:
\begin{equation} 
\log[\Pi_\omega(\phi_{\rm max}(\omega)-\varepsilon)]\sim
-\frac{\log^2(\varepsilon)}
{2\log(r)}, \quad {\rm with}\quad 
r=\frac{1-\phi_{\rm max}(\omega)}{\phi_{\rm max}(\omega)}.
\label{leading}
\end{equation}

\subsection{Tail of the distribution}

In fact, this complicated $ansatz$ was originally found by applying a similar
method as that of section {\bf III.C}, that we present now.

Again, let us define $P_\omega(\varepsilon)=
\int^{\phi_{\rm max}(\omega)}_{\phi_{\rm max}(\omega)-\varepsilon}
\Pi_\omega(\phi)\,d\phi$, which satisfies:
\begin{equation} 
P_\omega(\varepsilon)=\int_0^{1/4}dx\int_0^{\phi_{\rm max}(\omega)}d\phi\,
\sigma(x)\Pi_\omega(\phi)
\theta\left(\frac{x}{(1+\omega)^2(1-\phi)}+\varepsilon-\phi_{\rm max}(\omega)
\right).
\end{equation}
Once we express the actual domain of integration by imposing that the argument
of the $\theta$ function be positive, we find that the variable $x$ remains
very close to $1/4$ where the distribution $\sigma(x)$ is essentially
constant. Exploiting this fact, we find after a few elementary
manipulations that,
\begin{equation} 
P_\omega(\varepsilon)\sim C\int_0^{r\varepsilon}P_\omega
\left(u\right)\,du, 
\end{equation}
with $r=\frac{1-\phi_{\rm max}(\omega)}{\phi_{\rm max}(\omega)}>1$, and
$C=\frac{\sigma(1/4)}{4r\phi_{\rm max}(\omega)}$.  This equation is 
a rigorous integrated version of \eqn{(\ref{anz2})}. Note that for small
$\omega$, we have,
\begin{equation}
r=1+2\sqrt{2\omega}+O(\omega^{3/2}).
\label{eqr}
\end{equation}
We can now proceed in the same spirit as we did in section {\bf III.C},  and
find exact inequalities for $P_\omega(\varepsilon)$. For any $c_+>C$, and
sufficiently small $\varepsilon$:
\begin{eqnarray}
P_\omega(\varepsilon)&\leq c_+\int_{\varepsilon}^{r\varepsilon}P_\omega
(u)\,du+c_+\int^{\varepsilon}_0& P_\omega(u)\,du,\\
&\leq c_+(r-1)\varepsilon P_\omega(r\varepsilon)+c_+\varepsilon &
P_\omega(\varepsilon),
\end{eqnarray}
which finally leads, for small enough $\varepsilon$, to the existence of a
constant $C_+$ of order $O(1)$, such that,
\begin{equation} 
P_\omega(\varepsilon)\leq C_+(r-1)\varepsilon P_\omega(r\varepsilon).
\label{ineqg1}
\end{equation}
On the other hand, for any $r'<1$ close to 1, to be specified later, we can
write that,
\begin{eqnarray}
P_\omega(\varepsilon)& \geq C_-\int_{r'r\varepsilon}^{r\varepsilon}P_\omega
(u)\,du,&\\
& \geq C_-(1-r')\varepsilon P_\omega(r'&r\varepsilon),
\label{ineqg2}
\end{eqnarray} 
Let us now start from a small enough $\varepsilon_0=\varepsilon$ such that the
inequalities of  \eqn{(\ref{ineqg1},\ref{ineqg2})} hold, and define, 
\begin{eqnarray}
\varepsilon^+_{n}&=r^{-1}\varepsilon^+_{n-1}=r^{-n}\varepsilon,&\\
\varepsilon^-_{n}&=(r'r)^{-1}\varepsilon^-_{n-1}=(r'&r)^{-n}\varepsilon.
\end{eqnarray}
In the following, we will choose $r'$ such that $r'r>1$, so that
$\varepsilon^-_{n}\to 0$, when $n\to +\infty$.

By iterating the recursion inequalities we get,
\begin{eqnarray}
-\log(P_\omega(\varepsilon^+_n))&\geq -n\log(C_+(r-1))-&\sum_{k=0}^{n-1}\log
(\varepsilon^+_k)-\log(P_\omega(\varepsilon)),\\
& \sim -n\log(C_+(r-1))+&\frac{n^2}{2}\log(r),
\end{eqnarray}
where, after using $n\sim -\frac{\log(\varepsilon^+_n))}{\log(r)}$, 
the last line can also be written,
\begin{equation} 
\frac{\log^2(\varepsilon^+_n)}{2\log(r)}+
\frac{\log(C_+(r-1))}{\log(r)}\log(\varepsilon^+_n). 
\end{equation}
Similarly,
\begin{eqnarray}
-\log(P_\omega(\varepsilon^-_n))&\leq -n\log(C_-(1-r'))-\sum_{k=0}^{n-1}&\log
(\varepsilon^-_k)-\log(P_\omega(\varepsilon)),\\
& \sim -n\log(C_-(1-r'))+\frac{n^2}{2}\log&(r'r),
\end{eqnarray}
where the last line can also be written under the form,
\begin{equation} 
\frac{\log^2(\varepsilon^-_n)}{2\log(r'r)}+
\frac{\log(C_-(1-r'))}{\log(r'r)}\log(\varepsilon^-_n). 
\end{equation}
As $r'$ can be arbitrarily close to 1, we thus find that the leading order of
\eqn{(\ref{leading})} is exactly recovered.

Now, we restrict ourselves to the case of small $\omega$, and analyse the
effect of the subleading term. To be specific, let us take
$1-r'=\omega^{1/2+\delta}$, with $\delta$ arbitrarily small, such that the
condition $r'r>1$ remains satisfied (see \eqn{(\ref{eqr})}).
Our final results is that $P_\omega^-(\varepsilon)\leq P_\omega(\varepsilon)
\leq P_\omega^+(\varepsilon)$, with,
\begin{eqnarray}
-&\log(P_\omega^+(\varepsilon))&\sim
\frac{\log^2(\varepsilon)}{4\sqrt{2\omega}}+
\frac{\log(\omega)}{4\sqrt{2\omega}}\log(\varepsilon),\\
-&\log(P_\omega^-(\varepsilon))&\sim
\frac{\log^2(\varepsilon)}{4\sqrt{2\omega}}+
(1+2\delta)\frac{\log(\omega)}{4\sqrt{2\omega}}\log(\varepsilon),
\end{eqnarray}
which strongly suggests that,
\begin{eqnarray}
-\log(\Pi_\omega(\phi_{\rm max}(\omega)-\varepsilon))\sim
-\log(P_\omega(\varepsilon))\sim
\frac{\log(\varepsilon)\log(\omega\varepsilon)}{4\sqrt{2\omega}}.
\label{sing}
\end{eqnarray}

Note that the case $\omega=0$ falls exactly in the class of problem studied in
section {\bf III.C} leading to the exact asymptotics (with $\phi_{\rm
max}(0)=1/2$):
\begin{equation}
\log(\Pi_{\omega=0}(1/2-\varepsilon))\sim
\log(P_{\omega=0}(\varepsilon))\sim \frac{\log(\varepsilon)}{\varepsilon}.
\label{eqo0}
\end{equation} 

\subsection{Survival return probability}

The results of this section will not rely on as firm mathematical grounds as
that of the preceding sections, but will appear to be quite reasonable.

The comparison of \eqn{(\ref{sing})} and \eqn{(\ref{eqo0})} suggests that the
asymptotics of  \eqn{(\ref{sing})} should be correct at least up to
$\varepsilon$ of order $\sqrt{\omega}\sim 1/2-\phi_{\rm max}(\omega)$ (or more
exactly $\frac{\sqrt{\omega}}{|\log(\omega)|}$). Using \eqn{(\ref{survi})}, we
find that there should be a singular contribution in $\langle
p_{0,0}(\omega)\rangle$ of order (see \eqn{(\ref{sing})}),
\begin{equation}
\left[\langle p_{0,0}(0)\rangle-\langle p_{0,0}(\omega)\rangle
\right]_{\rm sing}\sim 
\int_{\phi_{\rm max}(\omega)}^{1/2}\Pi_{\omega=0}(x)\,dx\sim
\exp\left(\frac{c_1\log(\omega)}
{\sqrt{\omega}}\right),
\label{laplace}
\end{equation} 
where the last estimate comes from \eqn{(\ref{eqo0})}. In principle, the
leading singular correction to \eqn{(\ref{laplace})} should come from the
contribution of $\Pi_{\omega}$ to the integral of \eqn{(\ref{survi})} on  the
interval $[0,\phi_{\rm max}]$ and should be of order
$\exp\left(-\frac{c_2\log^2(\omega)}{\sqrt{\omega}}\right)$.

A contribution of the form $\exp(-c|\log(\omega)|^{\alpha'}/\omega^\alpha)$ in a
Laplace transform  generally originates from a large time decay of the form
(as found by a steepest descent type of argument),
\begin{equation}
\exp\left(-Ct^{\frac{\alpha}{1+\alpha}}\log^{\frac{\alpha'}{1+\alpha}}(t)\right).
\end{equation}  If we take the logarithmic corrections in
\eqn{(\ref{laplace})} seriously, we thus find:
\begin{equation}
\langle p_{0,0}(t)\rangle\sim\exp\left(-Ct^{1/3}\log^{2/3}(t)\right),
\label{decay}
\end{equation}
in disagreement with the heuristic argument given in \cite{gia2} (see 
\eqn{(\ref{ps})}).

\subsection{Exact bound for the survival return probability}

In this subsection, we give an exact lower bound for the number of surviving
particles using an argument which can be easily generalized to obtain a
similar bound for $\langle p_{0,0}(t)\rangle$. The resulting bound  is fully
consistent with \eqn{(\ref{decay})} and contradicts the heuristic estimate of
\cite{gia2} (see  \eqn{(\ref{ps})}).

Consider the symmetric model ($w_{i,i+1}=w_{i+1,i}$). The probability to have
a region of $L$ sites on which all $w_{1,2},\ldots,w_{L-2,L-1} >
1/2-\varepsilon$ is,
\begin{equation}
{\cal P}_L(\varepsilon)=\left[1-(1-2\varepsilon)^{1-\beta}\right]^{L-2}
\sim [2(1-\beta)\varepsilon]^L. 
\end{equation}
If we were interested in the non symmetric model, the corresponding
probability to have $w_{1,2},w_{2,1}\ldots,w_{L-2,L-1},w_{L-1,L-2} >
1/2-\varepsilon$ would be simply,
\begin{equation}
{\cal P}_L(\varepsilon)=\left[1-(1-2\varepsilon)^{1-\beta}\right]^{2(L-2)}
\sim [2(1-\beta)\varepsilon]^{2L}, 
\end{equation}
and the rest of the argument would be essentially identical. On such a region,
the density cannot decay faster than that of the following problem where we
consider $w_{1,2},\ldots,w_{L-1,L}=1/2-\varepsilon$, with fully absorbing
boundary conditions ($w_{0,1}=w_{1,0}=w_{L-1,L}=w_{L,L-1}=0$). This simple
property  can be shown by induction using the master equation for $p_i(t)$,
the density at site $i$: 
\begin{equation}
p_i(t+1)=w_{i+1,i}p_{i+1}(t)+w_{i-1,i}p_{i-1}(t),\quad p_i(t=0)=1.
\end{equation}
This simpler problem can be solved exactly (on the lattice or in the
continuum), leading to the following bound for the average decay of the total
density:
\begin{equation} 
n_L(t)>\frac{8}{\pi^2}\exp\left(-\frac{\pi^2t}{2L^2}-2\varepsilon t\right).
\label{nL}
\end{equation}
Finally, we find that the total density $n(t)$ is bounded for any $L$ and
$\varepsilon$ by the following exact lower bound:
\begin{equation} 
n(t)>C_0\exp\left(-\frac{\pi^2t}{2L^2}-2\varepsilon
t+L\log(2(1-\beta)\varepsilon)\right),
\end{equation}
where $C_0$ is a positive constant. We can now take the maximum of this lower
bound over $L$ and $\varepsilon$. Expressing this condition, we get the
following optimal values for $L$ and $\varepsilon$ defined implicitely by,
\begin{equation} 
L=2\varepsilon t=\left[-\frac{\pi^2 t}{\log(2(1-\beta)\varepsilon)}
\right]^{1/3}\sim \left[\frac{3\pi^2 t}{2\log(t)}\right]^{1/3},
\end{equation}
where the last estimate is valid for large time.

Finally, we have shown that,
\begin{equation} 
n(t)>\exp-S(t),\quad {\rm with}\quad S(t)\sim 
\left[\frac{3\pi^2 t}{2}\right]^{1/3}\log^{2/3}(t),
\end{equation}
in perfect agreement with the above analytical argument. The physical
interpretation of this result is that surviving particles are living in large
regions where the $w_{i,i\pm 1}$ are very close to $1/2$, and annihilates with
a large probability outside these regions. This explains why the symmetry of
the hopping probabilities is irrelevant and why the results is essentially
similar as that of perfectly diffusing particles with randomly distributed
perfect traps. The $\log^{2/3}(t)$ corrections is due to the fact that for
large time, the $w_{i,i\pm 1}$ in regions where the surviving particles stand
must be closer and closer to $1/2$, with an allowed fluctuation decreasing as
$\varepsilon\sim t^{-2/3}$. Moreover, within these large regions there is an
extra probability leaking per site of order $\varepsilon$.  Note finally, that
if $0<\gamma<1$, the argument can be repeated with $\varepsilon t$ replaced by
$(1-\gamma)\varepsilon t$ in \eqn{(\ref{nL})}, leading to the same decaying
behavior.

Let us now exhibit the flaw(s) in the argument given in \cite{gia2}, leading to 
\eqn{(\ref{ps})}. On a large region of size $L$ with $w_{i,i\pm
1}>1/2-\varepsilon$, the authors of \cite{gia2} estimated the probability
decay as,
\begin{equation} 
n_L(t)\sim\exp\left(-\frac{t}{L^{1/\nu}}-\varepsilon L^{1/\nu}\right),
\label{faux}
\end{equation}
where $\nu$ is the effective diffusion exponent defined in \eqn{(\ref{diffu})}
and below. This estimate is to be compared to our \eqn{(\ref{nL})}. The first
term is supposed to represent the probability decay due to the absorption of
particles at the boundaries of the considered region. It is not correct to
consider that particles in this region display anomalous diffusion as the 
$w_{i,i\pm 1}$ are in fact very close to $1/2$. This fact is confirmed by the
exact bound \eqn{(\ref{nL})}. The second term in the exponential is supposed
to reflect the fact that there is a small trapping probability (of order
$\varepsilon$) on each site of the considered region. The authors of
\cite{gia2} assumed that time can be replaced by $L^{1/\nu}$. This is
obviously wrong as the decay due to this small trapping probability
explicitly depends on the time spent in the region but not on its size.
Finally, the authors did not realize that  in the final expression that they 
obtained, $\varepsilon$ (and not only $L$) should also be treated as a
variational parameter.

Note that for intermediate times, we expect that the density should
decay as,
\begin{equation} 
n(t)\sim \exp(-cN(t))\sim \exp(-Ct^{1/2}),
\end{equation}
where $N(t)$ is the number of different sites visited by a random walker. This
phenomenon is also well-known for the Donsker-Varadhan problem, for which this
behavior can actually dominate the numerically accessible time regime
\cite{havlin,pstat}.

Finally, the generalization to higher dimensions of this model is
straightforward. On a periodic lattice of coordination number $z$, the
hopping probabilities are bounded by $z^{-1}$ and particles disappear if they
do not move. The above argument suggests that,
\begin{equation} 
n(t)\sim \exp\left(-Ct^{\frac{d}{d+2}}\log^{\frac{2}{d+2}}(t)\right),
\end{equation}

\section{Conclusion}

In this paper, we have considered a model where the trapping probabilities are
strongly correlated with the hopping probabilities of the walker. We have
shown that the escape probabilities have a well defined distribution which has
been analyzed in great detail in section {\bf III}. We have also explained
why this quantity display an apparent multifractal distribution. In section
{\bf IV}, we generalized our approach to the study of the survival return
probability distribution. We deduced from this exact analysis that this
survival probability (and the density) decays as
$\exp(-Ct^{1/3}\log^{2/3}(t))$. To support this result, we have obtained an
exact bound for the density which even reproduces the correct power of the
logarithmic correction. Moreover, we have explained the independence of the
model properties with respect to the disorder strength $\beta$, the trapping
rate $\gamma>0$ and, more surprisingly,  the symmetry of the hopping
probabilities.

A challenging problem is the understanding of the diffusion properties in this
model. The fact that the effective spreading  of the survivors is apparently
faster than diffusive (see \eqn{(\ref{diffu})}) remains to be explained.

\acknowledgements

I am grateful to F. van Wijland, C. Monthus and S. Redner for helpful
discussions on this problem, and D. Dean for useful comments on the manuscript.

\begin{figure}[ht]
\begin{center}
\includegraphics[angle=0,totalheight=9cm]{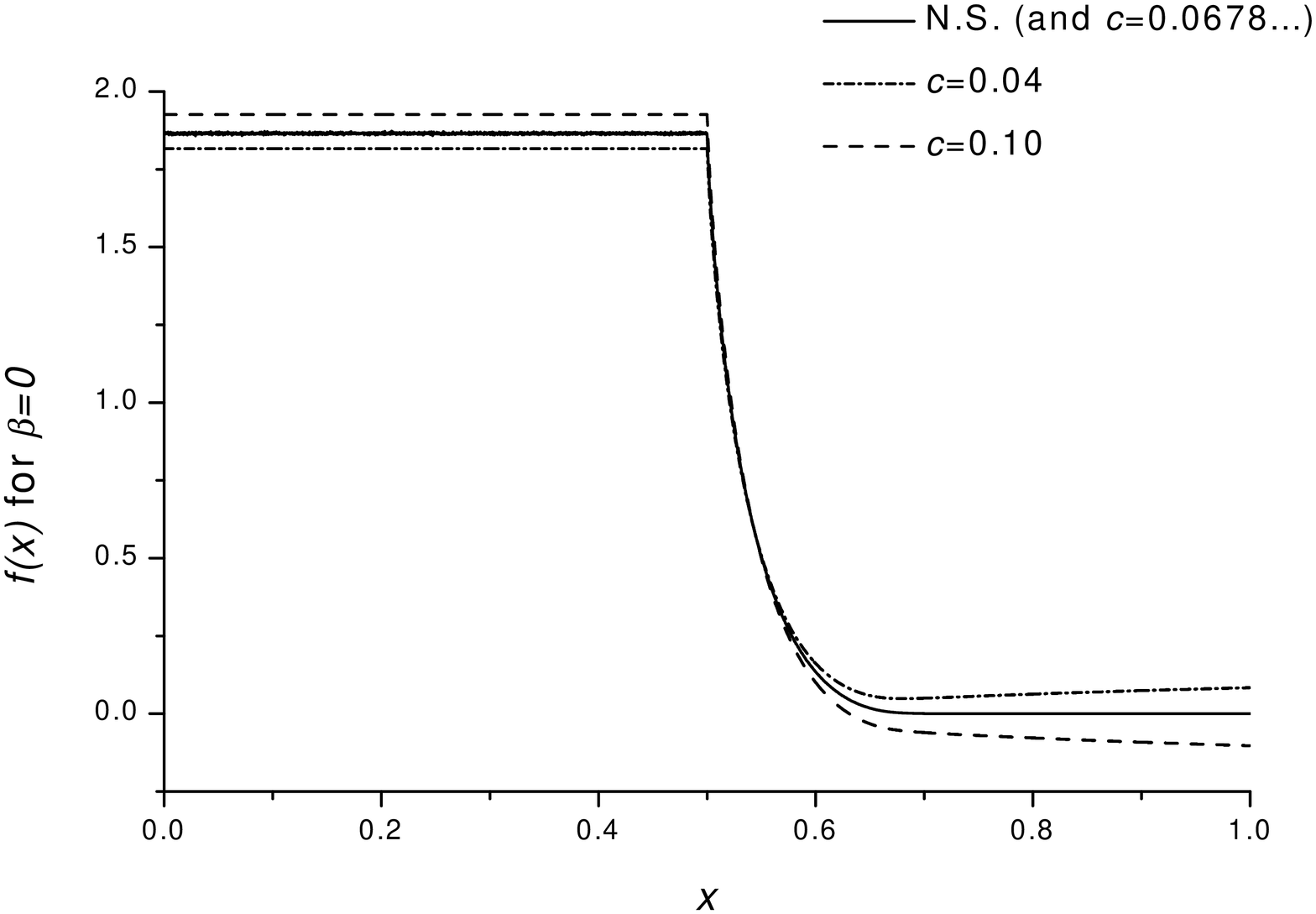}   
\caption{In the case $\beta=0$ ($f(x)=g(x)$), we plot the distribution 
obtained by iterating \eqn{(\ref{map})}  $2\times 10^9$ times (N.S.), and the
solution of \eqn{(\ref{diff})} obtained by imposing $f(1)=0$, which leads to
$c=0.0678\ldots$ (see text). The two curves are indiscernable. We also plot
the normalized solution of \eqn{(\ref{diff})} for $c=0.04$ and $c=0.10$.}
\end{center}
\end{figure}

\begin{figure}[ht]
\begin{center}
\includegraphics[angle=0,totalheight=9cm]{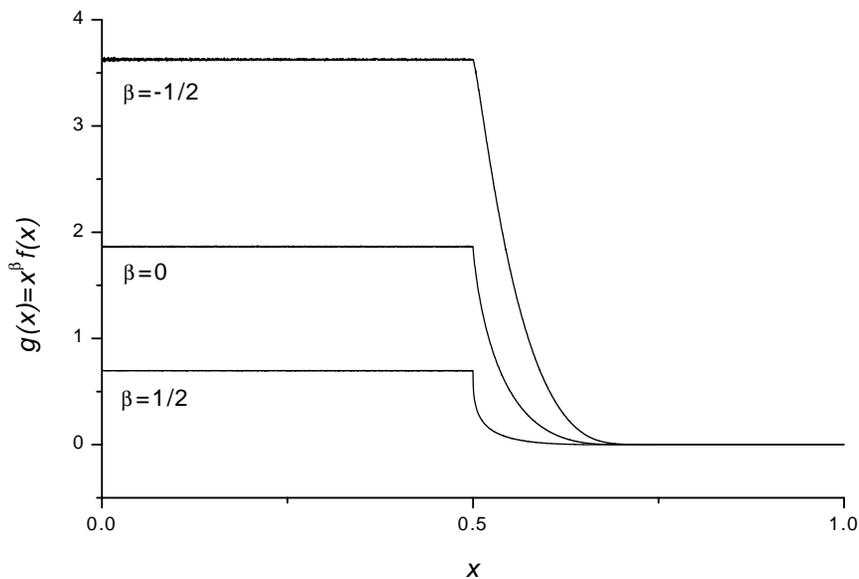}  
\caption{We plot $g(x)=x^\beta f(x)$ for $\beta=-1/2,0,1/2$. In each case, the
distribution is obtained by iterating $2\times 10^9$ times the map of
\eqn{(\ref{map})}. The agreement with the numerical solution of
\eqn{(\ref{diff})} is perfect.}
\end{center}
\end{figure}

\begin{figure}[ht]
\begin{center}
\includegraphics[angle=0,totalheight=9cm]{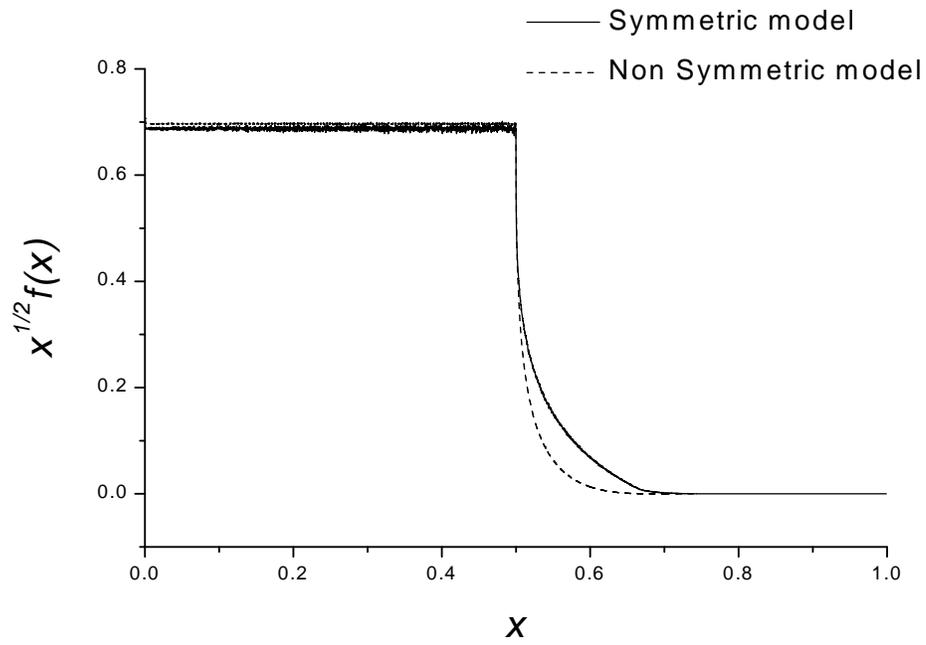}  
\caption{We plot $g(x)=x^\beta f(x)$, for $\beta=1/2$, and after $2\times 10^9$
iterations of the symmetric and non symmetric maps. As explained in the text,
the distribution decays faster in the non symmetric case.}
\end{center}
\end{figure}

\end{document}